\documentclass[draft]{agujournal2019}
\usepackage{url}
\usepackage{lineno}
\usepackage[inline]{trackchanges}
\usepackage{soul}

\usepackage{xcolor}

\newcommand{\aap}{Astronomy and Astrophysics}
\newcommand{\aapr}{Astronomy and Astrophysics Review}
\newcommand{\aj}{Astronomical Journal}
\newcommand{\apj}{Astrophysical Journal}
\newcommand{\apjl}{Astrophysical Journal Letters}
\newcommand{\apjs}{Astrophysical Journal Supplement}
\newcommand{\araa}{Annual Review of Astronomy and Astrophysics}
\newcommand{\grl}{Geophysics Research Letters}
\newcommand{\gca}{Geochimica Cosmochimica Acta }
\newcommand{\icarus}{Icarus}
\newcommand{\jgr}{Journal of Geophysics Research}
\newcommand{\mnras}{Monthly Notices of the Royal Astronomical Society}
\newcommand{\nat}{Nature}
\newcommand{\pasp}{Publications of the Astronomical Society of the Pacific}
\newcommand{\planss}{Planetary Space Science}
\newcommand{\prl}{Physical Review Letters}
\newcommand{\ssr}{Space Science Reviews}

\draftfalse

\journalname{JGR: Planets}

\begin{document}

\title{The Fundamental Connections Between the Solar System and
  Exoplanetary Science}

\authors{
  Stephen R. Kane\affil{1},
  Giada N. Arney\affil{2},
  Paul K. Byrne\affil{3},
  Paul A. Dalba\affil{1}\thanks{NSF Astronomy and Astrophysics
    Postdoctoral Fellow},
  Steven J. Desch\affil{4},
  Jonti Horner\affil{5},
  Noam R. Izenberg\affil{6},
  Kathleen E. Mandt\affil{6},
  Victoria S. Meadows\affil{7},
  Lynnae C. Quick\affil{8}
}

\affiliation{1}{Department of Earth and Planetary Sciences, University
  of California, Riverside, CA 92521, USA}
\affiliation{2}{Planetary Systems Laboratory, NASA Goddard Space Flight
  Center, Greenbelt, MD 20771, USA}
\affiliation{3}{Planetary Research Group, Department of Marine, Earth,
  and Atmospheric Sciences, North Carolina State University, Raleigh,
  NC 27695, USA}
\affiliation{4}{School of Earth and Space Exploration, Arizona State
  University, Tempe, AZ 85287, USA}
\affiliation{5}{Centre for Astrophysics, University of Southern
  Queensland, Toowoomba, QLD 4350, Australia}
\affiliation{6}{Johns Hopkins University Applied Physics Laboratory,
  Laurel, MD 20723, USA}
\affiliation{7}{Department of Astronomy, University of Washington,
  Seattle, WA 98195, USA}
\affiliation{8}{Planetary Geology, Geophysics and Geochemistry
  Laboratory, NASA Goddard Space Flight Center, Greenbelt, MD 20771,
  USA}

\correspondingauthor{Stephen R. Kane}{skane@ucr.edu}

\begin{keypoints}
\item Exoplanetary science is rapidly expanding towards
  characterization of atmospheres and interiors.
\item Planetary science has similarly undergone rapid expansion of
  understanding planetary processes and evolution.
\item Effective studies of exoplanets require models and in-situ data
  derived from planetary science observations and exploration.
\end{keypoints}


\begin{abstract}

Over the past several decades, thousands of planets have been
discovered outside of our Solar System. These planets exhibit enormous
diversity, and their large numbers provide a statistical opportunity
to place our Solar System within the broader context of planetary
structure, atmospheres, architectures, formation, and
evolution. Meanwhile, the field of exoplanetary science is rapidly
forging onward towards a goal of atmospheric characterization,
inferring surface conditions and interiors, and assessing the
potential for habitability. However, the interpretation of exoplanet
data requires the development and validation of exoplanet models that
depend on in-situ data that, in the foreseeable future, are only
obtainable from our Solar System. Thus, planetary and exoplanetary
science would both greatly benefit from a symbiotic relationship with
a two-way flow of information. Here, we describe the critical lessons
and outstanding questions from planetary science, the study of which
are essential for addressing fundamental aspects for a variety of
exoplanetary topics. We outline these lessons and questions for the
major categories of Solar System bodies, including the terrestrial
planets, the giant planets, moons, and minor bodies. We provide a
discussion of how many of these planetary science issues may be
translated into exoplanet observables that will yield critical insight
into current and future exoplanet discoveries.

\end{abstract}


\section*{Plain Language Summary}
Thousands of planets have been found outside of our Solar System,
called "exoplanets", forging a new frontier of planetary
exploration. However, studying these planets many light years away
requires a deep understanding of the planets nearby so that we can
accurately interpret the planetary processes that are occurring on
these distant worlds. In this work, we provide a summary of advances
in planetary science and describe how the various Solar System bodies
enable us to unlock the secrets of exoplanets. These advances include
new insights into planetary habitability, and we discuss how
diagnosing the evolution of our nearest neighbors can further the
search for life in the universe.


\section{Introduction}
\label{intro}

Underpinning planetary science is a deep history of observation, and
more recently, planetary exploration within the Solar System, from
which models of planetary processes have been constructed
\cite<e.g.>[and references
  therein]{depater2015a,horner2020b}. Indeed, planetary science as a
discipline has greatly benefited from the robotic exploration of the
Solar System over the past 60 years. From the early 1960s onwards, we
began to explore beyond the Earth-Moon system, with flybys of Venus
and Mars \cite<e.g.>{fjeldbo1966b,neugebauer1966} followed,
eventually, by landings on those planets
\cite<e.g.>{avduevski1971b,keldysh1977,hess1977b,toulmin1977b}. At
the present time, we have now sent spacecraft past each of the eight
planets \cite<e.g.>{smith1982,smith1989}; have delivered orbiters
to each of the terrestrial planets
\cite<e.g.>{solomon2001,nakamura2011}, as well as the giants
Jupiter and Saturn \cite<e.g.>{belton1996c,spencer2006a}; and have
also visited several of the Solar System's dwarf planets
\cite<e.g.>{russell2011b,stern2015d} and smaller bodies
\cite<e.g.>{krankowsky1986,fujiwara2006,glassmeier2007a}. The
  detailed observations and in-situ measurements of Solar System
  bodies provide the basis of fundamental models that describe the
  origin and evolution of planetary systems, as well as the nature of
  atmospheric and geological planetary processes
  \cite<e.g.>{goldreich1966a,guillot1999b,lodders2003,zahnle2003a}.

\begin{figure}
  \begin{center}
      \includegraphics[width=8.5cm]{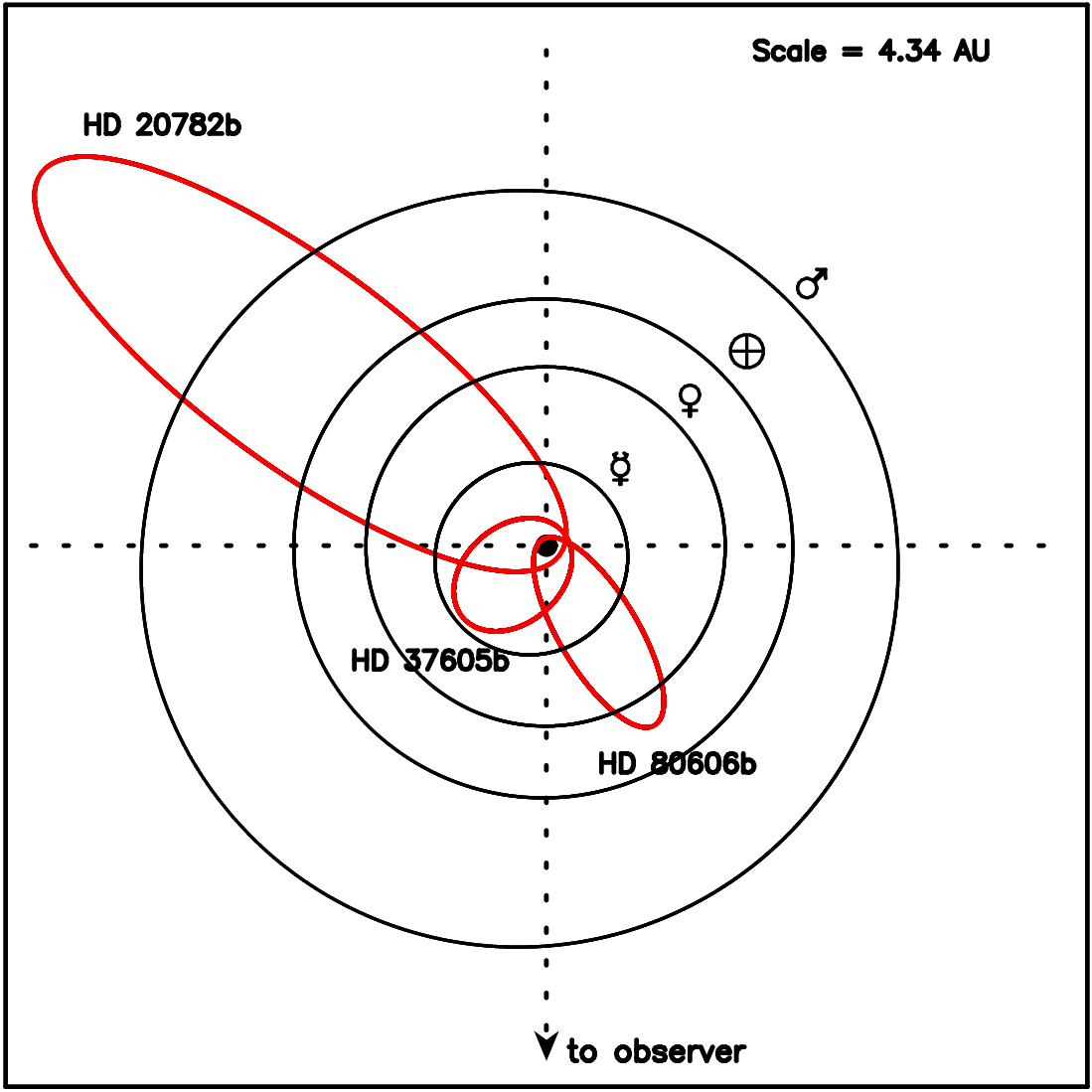}
  \end{center}
  \caption{Example orbits of highly eccentric exoplanets (shown in
    red); HD~20782b, HD~80606b, and HD~37605b. These orbits are
    overlaid on the orbits of the Solar System terrestrial planets,
    shown in black. The scale of the figure is 4.34~AU along a single
    side.}
  \label{fig:orbits}
\end{figure}

In parallel, the past few decades have seen the rapid expansion of
exoplanetary science. At present, the number of known exoplanets has
passed
4,300\footnote{\url{https://exoplanetarchive.ipac.caltech.edu/}}
\cite{akeson2013}. The current exoplanet inventory contains planets of
types vastly different to those in the Solar System, such as
super-Earths \cite{valencia2007b,leger2009,howard2010b,bonfils2013a},
mini-Neptunes \cite{barnes2009a,lopez2014,nielsen2020b}, and hot
Jupiters \cite{mayor1995,fortney2008a,wright2012}. Furthermore,
notable demographic trends have been detected, such as the deficit in
planets with radii between 1.5--2.0 Earth radii with orbital periods
less than 100 days
\cite<e.g.>{fulton2017,fulton2018b,vaneylen2018b,berger2020b}.
Multi-planet systems feature an enormous diversity of architectures
\cite{ford2014,winn2015,hatzes2016d,he2019}, including compact
systems, such as Kepler-11 \cite{lissauer2011a}, that demonstrate the
capacity of planetary systems to harbor multiple planets interior to a
Mercury-equivalent orbit. Furthermore, exoplanetary systems display a
broad range of orbital eccentricities compared with the near-circular
orbits of the planets in the Solar System. Shown in
Figure~\ref{fig:orbits} are the orbits for the exoplanets HD~20782b
\cite{kane2016b}, HD~80606b \cite{bonomo2017}, and HD~37605b
\cite{wang2012}, overlaid on the orbits of Solar System planets. A
complete analysis of exoplanetary statistics offers important insight
into the question of how typical our Solar System architecture and
evolution is \cite{limbach2015,martin2015b}. Furthermore, future
observations of terrestrial exoplanet atmospheres have the promise to
transform our understanding of the rocky planets in the Solar System
\cite{shields2019a}, with particular astrobiological significance for
the study of planets within the Habitable Zone (HZ) of their stars
\cite{kasting1993a,kane2012a,kopparapu2013a,kopparapu2014,kane2016c}.

There are substantial challenges facing an efficient and seamless
integration of planetary and exoplanetary science, however, largely
related to the language, techniques, and measurables that are
prevalent in these respective fields. For instance, planetary science
directly studies the atmosphere, geology, and interiors of planets for
which we have spatially and temporally resolved and/or in-situ
data. Yet, at the present time, our knowledge of exoplanet properties
is usually not directly obtained since the planets remain invisible to
us, and is instead inferred from the planet's impact on the host
star's orbit or brightness. Knowledge of stellar astronomy then
becomes the baseline needed to understand planetary
characteristics. Nonetheless, it is clear that the boundaries between
these two fields, including language, terminology, methodology, and
sharing of results/data, are worth dismantling if a full understanding
of planets at the systems level is to be realized. Exoplanetary
science provides a statistical insight into planetary architectures
and formation scenarios, and planetary science provides detailed
planetary models that exoplanetary science relies upon for detailed
characterization. For example, a detailed study of the processes
governing past and current atmospheric escape for the Solar System
terrestrial planets, giant planet moons, and small bodies plays an
important role in understanding the survival and evolution of
exoplanet atmospheres
\cite{strangeway2005,tian2015b,dong2018c,gronoff2020b,lammer2020b}.

There are numerous reasons why the study of Solar System planets and
exoplanets in unison is critical for the advancements of both fields,
including:
\begin{enumerate}
\item Terrestrial exoplanets are extremely common \cite{winn2015},
  and will form the basis for a large-scale effort towards
  measurements of planetary atmospheric characteristics
  \cite{kempton2018,lustigyaeger2019a}, which will, in turn, be
    applied to understanding Solar System atmospheric abundances
    \cite{martin2015b,bean2017}.
\item Studies have shown that giant planets drive the architecture and
  evolution of planetary systems
  \cite{gomes2005b,morbidelli2005,walsh2011c,raymond2014a,nesvorny2018c},
  and may play a major role in water delivery to terrestrial planets
  \cite{obrien2014a}.
\item Planetary (Solar System) science is continually advancing, with
  frequent and often considerable revisions to our understanding of
  fundamental processes, and to prevailing models of formation,
  dynamics, atmospheres, surfaces, and interiors
  \cite<e.g.>{lodders2003,tsiganis2005b,adams2010b,mitchell2016a,wordsworth2016b,lunine2017,read2018b}.
\item Perhaps most importantly is the knowledge that we will not have
  access to in-situ data for an exoplanet within the foreseeable
  future, such that exoplanet surface conditions will predominantly be
  inferred from models based on Solar System data.
\end{enumerate}

In this paper, we present a summary of the major bodies within the
Solar System, their interiors and atmospheres, major outstanding
questions, and the relevance of these worlds to exoplanetary
science. In Section~\ref{terrestrial} we outline the properties of the
terrestrial planets, in Section~\ref{giant} we describe the gas and
ice giant planets, in Section~\ref{moons} we discuss the relevant
properties of major icy moons in the Solar System, and in
Section~\ref{minor} we provide the lessons learned from minor
planets. Section~\ref{exoplanets} describes the progress in
exoplanetary science, and how discoveries made for worlds in this
planetary system will contribute to the interpretation of data for
others in the coming years. We provide a brief summary for the overlap
between planetary and exoplanetary science along with concluding
remarks in Section~\ref{conclusions}.


\section{The Terrestrial Planets}
\label{terrestrial}

\begin{figure*}
  \begin{center}
      \includegraphics[width=\textwidth]{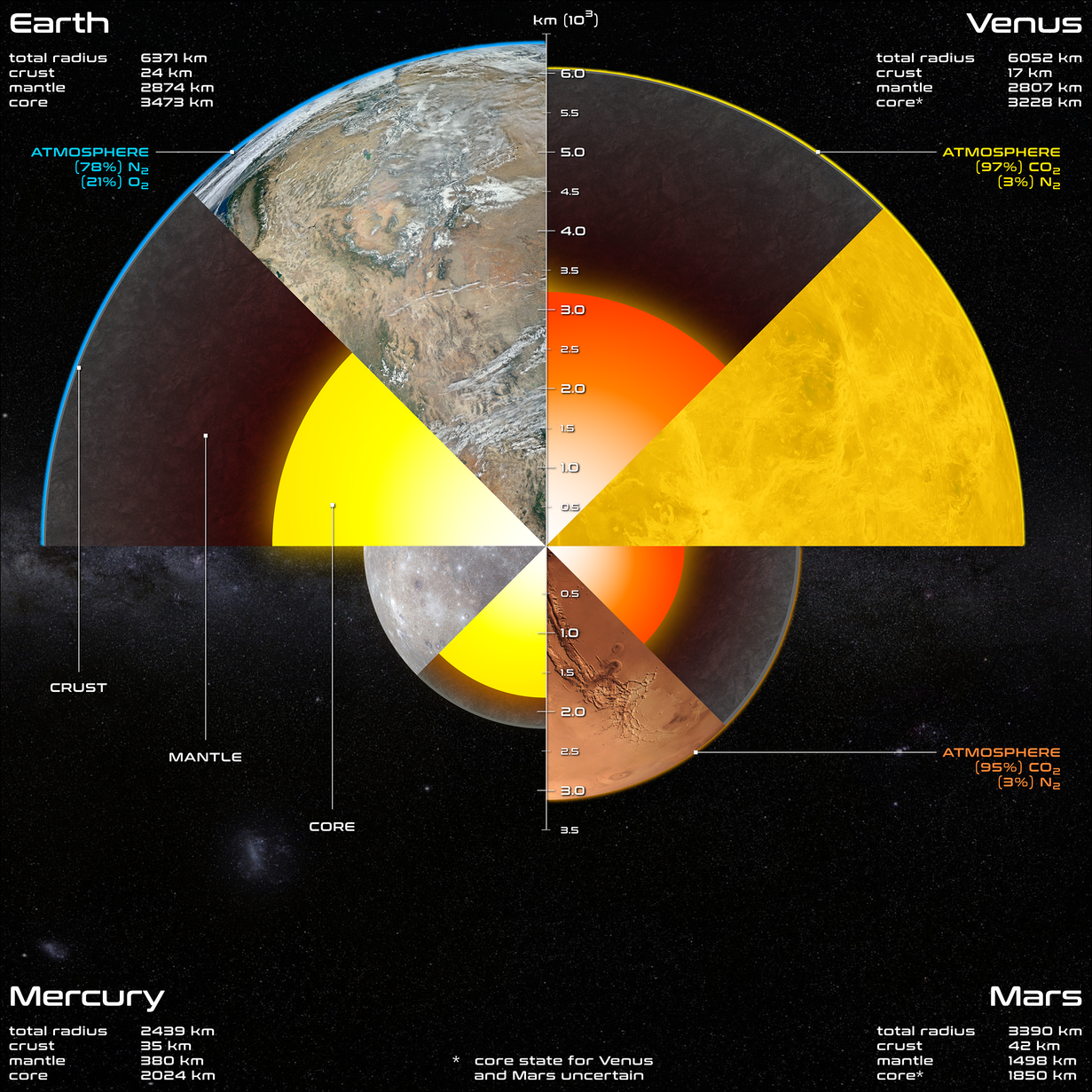}
  \end{center}
  \caption{Schematic cross sections of the four inner Solar System
    planets, showing the major internal components (crust, mantle, and
    core) and atmospheric components. All cross sections are to
    scale. For simplicity, oceanic and continental crust for Earth are
    not distinguished, nor is the interior structure of Earth's mantle
    shown. Note that there is considerable uncertainty regarding the
    state of the cores of Venus and Mars, and specifically whether
    there is a liquid and solid component, as for Earth's core, and so
    they are shown with orange fill instead of yellow. The interior
    structure for Earth is from \citeA{dziewonski1981}. For Venus, the
    crust–mantle depth and mantle–core depth values are from
    \citeA{james2013} and \citeA{aitta2012}, respectively. For Mars,
    those values are from \citeA{goossens2017} and \citeA{plesa2018b},
    respectively, and for Mercury are from \citeA{padovan2015} and
    \citeA{smith2012e}, respectively. The exosphere of Mercury is not
    shown.}
  \label{fig:terrestrial}
\end{figure*}

The terrestrial planets of the Solar System serve as a foundation for
our understanding of rocky planet interiors, atmospheres, and
evolution generally. Shown in Figure~\ref{fig:terrestrial} are
schematic cross-sections for the four Solar System terrestrial
planets, used here to illustrate their relative sizes and
known or inferred interior structure. In particular,
characterizing the conditions and properties of these worlds help
us develop models with which to understand how surface conditions
can reach equilibrium states that are temperate and potentially
habitable, or hostile with thick and/or eroded atmospheres. Here, we
summarize the primary features of the terrestrial planets, and some of
the outstanding questions that remain regarding their properties.


\subsection{Mercury}
\label{mercury}

The orbital reconnaissance of the inner Solar System planets was
completed by observations returned by the {\it MESSENGER}
spacecraft. Mercury is a world that experienced sustained, widespread
effusive volcanism for the first quarter of its life, before interior
cooling and global contraction outpaced melt production and the
prevailing stress state shut off major volcanic activity around 3.5~Ga
\cite{solomonbyrne2018}. {\it MESSENGER} saw a world heavily scarred
by impact bombardment but also surprisingly volatile rich for a rocky
planet so close to its parent star-with the planet having an unusually
high abundances of S and C \cite{vanderkaaden2017} and even evidence
for ongoing sublimation of a volatile-rich crust
\cite{solomonblewett2018}. (Note that we do \textit{not} include
"water" here as a volatile, which is often a major volatile species in
discussions of planetary formation, because the water content of
Mercury is poorly constrained.) How a planet in such proximity to the
Sun could retain relatively high volatile abundances remains a key
outstanding question.

Equally surprising was the discovery with {\it MESSENGER} data of a
remarkably thin silicate portion for Mercury: the core-mantle boundary
is only 420 km deep. One possibility is a formation scenario whereby
Mercury naturally accreted with much more iron, and much less silica,
than the other inner Solar System bodies—although this interpretation
presents some substantial chemical challenges
\cite{solomon2018}. Alternatively, a single, catastrophic collision
\cite{benz2007} or several "hit-and-run" impacts
\cite{asphaug2014b,jackson2018a} may have stripped off much of the
outer portion of an originally larger "proto-Mercury". The prospect of
a major impact shaping Mercury is supported by evidence for other
similar events in early Solar System history, including the formation
of the Earth-Moon system. If impact stripping of a proto-Mercury is
true, then such dramatic reshaping of the planet happened very early
in its history; the oldest preserved terrain on the planet is ~4.1 Ga
\cite{marchi2013b}, with most of the earlier history of the planet
likely buried by major volcanic and impact ejecta deposits
\cite{solomon2018}. Although impact stripping may deplete a planet's
volatile inventory, and thus seem incompatible with Mercury's apparent
enrichment in moderately volatile elements such as S and C, impact
modeling suggests that such volatiles might be retained in a vapor
cloud to later recondense on the planet \cite<e.g.>{ebel2017}.

The relevance of Mercury to exoplanetary science lies in both its
chemical make-up and its interior structure: how can a rocky world
with high volatile abundance and an outsize core form so close to its
star? It may be that such compositions are possible in the inner
portions of the protoplanetary disk. Or, perhaps, giant impacts are
not all that unusual. For example, \citeA{bonomo2019} noted that the
planets Kepler-107b and Kepler-107c have almost identical diameters
(both $\sim$1.5--1.6 $R_\oplus$), but significantly divergent
densities ($\sim$12.6~gcm$^{-3}$ for Kepler-107c
vs. $\sim$5.3~gcm$^{-3}$ for Kepler-107b). Improving our knowledge of
how Mercury came to form with its present interior structure will
provide valuable information with which to assess which of these
scenarios—initial accretion, or subsequent impact-stripping—may apply
to Kepler-107b.

Some key questions of Mercury relevant to exoplanetary science
include (but are by no means limited to):
\begin{enumerate}
\item Did Mercury form with its large core, or did giant impact(s)
  strip away much of its silicate material
  \cite<e.g.>{benz2007,ebel2017}?
\item What effect, if any, has the proximity of Mercury to the
  host star had on the planet's abundances of moderately volatile
  elements such as C and S \cite<e.g.>{vanderkaaden2020}?
\item What does the current surface geology tell us of the interior
  structure and thermal history of a planet with such a large core and
  relatively modest mantle mass fraction
  \cite<e.g.>{solomonbyrne2018}?
\item How has the composition and geology of Mercury’s airless surface
  being affected by its close proximity to its host star
  \cite<e.g.>{solomonchapman2018}?
\end{enumerate}

In terms of exoplanet science, perhaps more than anything else Mercury
offers us a natural laboratory for understanding how a rocky planet
close to its star can form with—and retain over geological
time—substantial inventories of moderately volatile species such as C
and S, which might have been expected to be absent given the planet's
distance to the Sun. Additionally, determining how Mercury attained
its outsize core will in turn tell us what role, if any, giant impacts
play in the formation of such planetary interior structure. Detecting
and observing small rocky planets close to their host stars remains
technically challenging, but Mercury is a useful basis with which to
interpret similarly sized and structured close-in worlds elsewhere.


\subsection{Venus}
\label{venus}

Venus is often referred to as Earth's "sibling", because it is the
Solar System world most alike to Earth in terms of size and bulk
composition. However, it has a $\sim$92-bar atmosphere comprising
96.5\% CO$_2$ and 3.5\% N$_2$, and a surface temperature of
$\sim$735~K. The surface of Venus has a surprisingly young average age
of $\sim$750~Ma, based on crater counts \cite<e.g.>{schaber1992}, and
may be in a "stagnant lid" state \cite<e.g.>{herrick1994b}, although
subduction may still occur today \cite{davaille2017,smrekar2018}. The
standard explanation for the present atmospheric state of Venus is a
past progression into a runaway greenhouse \cite{walker1975} that
occurred when incident solar radiation (insolation) exceeded the limit
on outgoing thermal radiation from a moist atmosphere
\cite{komabayasi1967,ingersoll1969c,nakajima1992,goldblatt2012},
evaporating any oceans present. The possibility of past oceans on the
surface of Venus is not a new concept
\cite<e.g.>{kasting1984a,kasting1988c,grinspoon1993}. However,
recently \citeA{way2020} suggested an alternative scenario for surface
water rentention, in which ocean evaporation was not dominated by
effects related to secular changes in insolation, based on cloud decks
forming at the sub-stellar point. Instead, these authors argued, the
emission into the atmosphere of amounts of CO$_2$ greater than could
be drawn down into the surface and a putative ocean, most likely by
major volcanic eruptions, would have triggered a moist greenhouse
scenario. If so, then the fundamental difference in climate between
Venus and Earth may be more stochastic, and less inevitable, than
previously assumed. The explanation for this difference relies upon a
deeper understanding of the relative contributions of CO$_2$
outgassing compared with the physical properties of clouds that
determine albedo.

In any case, once a moist greenhouse effect was underway, it is likely
that water loss by hydrogen escape followed, evident in high D/H
relative to Earth \cite{donahue1982,debergh1991}. Complete water loss
for Earth-equivalent oceans would take a few hundred million years
\cite{watson1981,kasting1984a,zahnle1986b,kasting1988c}, depending on
extreme ultraviolet (XUV) flux and potential throttled by oxygen
accumulation \cite{wordsworth2013b}. Moreover, the Venusian nitrogen
inventory is poorly known and may hold important clues to the
atmospheric and mantle redox evolution
\cite{wordsworth2016a}. Notably, massive water loss during a moist and
runaway greenhouse has been suggested as producing substantial O$_2$
in exoplanet atmospheres \cite{wordsworth2014,luger2015b}, but the
present Venus atmosphere does not show this tracer of ocean loss and
potential false positive for an oxygen biosignature. Hydration and
oxidation of surface rocks \cite<e.g.>{matsui1986a} and
top-of-the-atmosphere loss processes
\cite{chassefiere1997,collinson2016a} may have removed any O$_2$ that
was produced by early ocean loss, although it is uncertain how the
present atmospheric loss processes would operate for a different
(younger) Venus atmosphere, potentially in the presence of a stronger
magnetic field \cite{luhmann2008b,curry2015a,persson2020}. Thus, Venus
is an ideal laboratory to test hypotheses for abiotic oxygen
production and loss processes.

The evolutionary history of Earth's sibling is of crucial importance
to not only understanding both the past and future of our world, but
the analysis of terrestrial exoplanet atmospheres more generally
\cite{schaefer2011,ehrenreich2012a,lustigyaeger2019b}. This
consideration is particularly important in the current era of
exoplanet detection, from which the discovery of potential Venus
analogs are expected to become the prime targets for detailed
follow-up observations \cite{kane2013d,kane2014e,ostberg2019}. An
example of a potential exoplanet analog to Venus is Kepler-1649b
\cite{angelo2017a}, for which climate simulations predict rapid
water-loss scenarios and possible progression through a runaway
greenhouse \cite{kane2018d}. However, the relative lack of knowledge
regarding the dynamics and chemistry of middle and deep atmosphere of
Venus presents a barrier to detailed modeling of surface environments
of terrestrial exoplanets \cite<e.g.>{forget2013,forget2014}. Thus,
Venus is a particularly important object of study within the Solar
System as a possible template for the expected challenges in
characterizing the evolution and atmospheres of terrestrial exoplanets
in or near the HZ \cite{kane2019d,lustigyaeger2019b}. Examples of
outstanding questions regarding Venus are:
\begin{enumerate}
\item What is the interior structure and composition of Venus
  \cite{gillmann2014,gulcher2020b,orourke2020}?
\item What has been the history of tectonics, volatile cycling, and
  volcanic resurfacing \cite{ivanov2011b}? Was the delivery of
  volatiles to the atmosphere from the surface and interior gradual,
  episodic, or catastrophic?
\item What is the detailed composition and atmospheric chemistry that
  exists within the Venusian middle and lower atmosphere
  \cite{krasnopolsky2012a,bierson2020}, and how does the lower
  atmosphere interact with the surface?
\item What drives the Venus atmospheric circulation
  \cite{fukuhara2017a,horinouchi2017}, and in particular the
  super-rotation on this slowly rotating planet
  \cite{lebonnois2010,horinouchi2020,lebonnois2020b}? How can the
  atmospheric dynamics of Venus be used to model tidally-lock
  exoplanets \cite<e.g.>{heng2011a,yang2019a}?
\item Where did Venus' water go, and what processes are most important
  for O$_2$ loss from terrestrial planet atmospheres? Was hydrogen
  loss and abiotic oxygen production prevalent, or did surface
  hydration dominate \cite{kasting1984a,watson1984,lichtenegger2016}?
\item Did Venus have a habitable period \cite{way2016,way2020}? That
  is, did Venus ever cool after formation \cite{hamano2013}? If Venus
  had a habitable period, how long did it last—and when did it end?
\item What is the nature of the unknown UV absorber in the Venus
  atmosphere \cite{esposito1980b,molaverdikhani2012,perezhoyos2018},
  responsible for absorbing half of the insolation into Venus'
  atmosphere, and could it have astrobiological significance for Venus
  and exoplanets \cite{limaye2018b}?
\end{enumerate}

Many of the remaining questions regarding Venus have strong overlap
with the community goals of understanding the evolution of
exoplanets. For example, the nature of water delivery to Venus remains
uncertain \cite{gillmann2020}, a factor that determines long-term
habitability \cite{way2016}. Additionally, atmospheric mass-loss
\cite<e.g.>{howe2020,kane2020d} and water-loss from the top of the
atmosphere \cite{wordsworth2013b} depend on XUV flux from star, which
in turn depends on spectral type \cite{dong2018a,owen2019a}. Moreover,
the relative lack of knowledge regarding the bulk composition of Venus
makes it difficult to infer the mineralogy of exoplanets based on
stellar abundances \cite{hinkel2018}. Most importantly, the evolution
of Venus potentially represents a pathway from habitable to
uninhabitable conditions, a pathway whose nature may be common for
terrestrial planets \cite{foley2015,foley2016,way2020}. Thus, the
study of planetary habitability will benefit from understand which of
the myriad of differences between Venus and Earth dominated the
divergence in their planetary evolutions \cite{kane2019d}.


\subsection{Earth}
\label{earth}

In the discussion of life beyond the Solar System, a great majority of
the focus lies on finding exoplanets that may be similarly habitable
to Earth. In many ways, it makes sense to focus our efforts on planets
of a similar size, mass, and insolation flux to Earth, because Earth
is the only known globally habitable and inhabited planet. For that
reason, studies of Earth are of critical importance in shaping the
future direction of exoplanetary science
\cite<e.g.>{horner2010e,unterborn2016,fan2019b,groot2020}. One of the
great challenges that such a focus on Earth-similar planets poses is
the determination of which factors in Earth's characteristics and
history are universally required for habitability and life
\cite{meadows2018a}. It is natural to look at Earth and ascribe our
existence to any and all of our planet's peculiar and unique features,
from the presence of our anomalously large satellite, to its
internally generated magnetic field and magnetosphere, to the
relatively benign impact regime our planet has experienced, at least
for the past few billion years. It is worth noting that ascribing
known life to fundamental Earth properties may, in some cases, present
an erroneous line of reasoning that requires further investigation to
properly resolve.

There have been numerous studies regarding the role of the Moon in
stabilizing the spin axis of the Earth
\cite<e.g.>{laskar1993b,lissauer2012a,barnes2016a,cuk2016b}, including
suggestions that the such stabilization may have moderated the Earth's
climatic variability \cite<e.g.>{waltham2004}, and therefore
habitability
\cite{williams1997b,spiegel2009a,heller2011a,armstrong2014b,colose2019}.
As such, the possible requirement of the presence of a substantial
moon for long-term habitability continues to be used as an argument
towards the potential scarcity of habitable planets in the
Universe. The reason for that assertion is that the formation of the
Moon is thought to have been a stochastic event, the result of a giant
collision between ``proto-Earth'' and a Mars-sized object
(colloquially referred to as ``Theia''
\cite<e.g.>{benz1986,canup2001b,reufer2012}). However, such stochastic
events does not necessarily mean that analog Earth-moon systems are
rare \cite{elser2011}. Additionally, it has been demonstrated that the
Earth may possibly maintain long-term obliquity stability without the
presence of the Moon \cite{lissauer2012a,li2014b}, reducing the
dependence of climate evolution on its presence.

Similarly, the origin of the Earth's volatile budget is still a point
of some discussion
\cite{marty2012,marty2016,dauphas2017,peslier2017,wu2018}. Theories
for the hydration of Earth fall into three broad groups:
\textit{endogenous hydration}, in which the water was accreted from
material local to the Earth from the solar neula \cite{ikoma2006a},
typically in the form of hydrated silicates
\cite<e.g.>{drake2005}; \textit{early exogenous hydration}, where
volatile material was delivered from beyond the ``ice-line'' as Earth
was still accreting, in the form of asteroids and comets flung inward
by the giant planets (possibly as the latter migrated)
\cite<e.g.>{morbidelli2000,petit2001b}; and \textit{late exogenous
  hydration} (otherwise known as the ``late-veneer'' family of
models), which invokes the delivery of water from the outer Solar
System towards the end of Earth's accretion, or even some time after
the formation of the planet was essentially complete
\cite<e.g.>{owen1995b}.

A common feature of many of these models is the implicit assumption
that all of the terrestrial planets received similar amounts of
volatile material, and that the isotopic abundances of the volatiles
delivered to them ought to have been the same from one planet to the
next. However, dynamical studies have shown that the different
terrestrial planets likely received different amounts of material from
different reservoirs of volatiles -- at least in the case of the
exogenous delivery of those volatiles
\cite<e.g.>{owen1996a,raymond2004a,horner2009b}, a result that has
been replicated in studies of planet formation around other stars
\cite<e.g.>{ciesla2015b}. This finding has implications, for
example, for the volatile inventory of Venus and its similarity (or
not) with that of Earth \cite<e.g.>{way2020}.

The composition of Earth's earliest atmosphere is poorly known,
although life may have evolved during the earliest phase of Earth
history, the Hadean ($>$4.0 billion years ago)
\cite{nutman2016}. Since life arose, atmospheric abundances of
biosignature gases (e.g., O$_2$, O$_3$, CH$_4$, N$_2$O) have varied by
orders of magnitude over our planet's billion year history
\cite{zahnle2007,schwieterman2018,zahnle2020}, with major implications
for the detectability and interpretation of the presence of these
gases on exoplanets. These changes in atmospheric composition have
also featured in the most dramatic changes in Earth overall
environmental history. In the Archean eon (4.0--2.5 billion years
ago), Earth's atmosphere is thought to have been relatively anoxic
\cite{lyons2014a}, though evidence exists for an earlier oxygen-rich
atmosphere \cite{ohmoto2020}. Because the Sun then was only 70--80\%
as luminous as today, enhanced greenhouse warming was necessary, and
perhaps sufficient to keep Earth clement during this time
period. These greenhouse gases likely included carbon dioxide (CO$_2$)
and methane (CH$_4$), and when present together in large quantities,
can indicate a biological atmospheric disequilibrium
\cite{krissansentotton2018a}. Methane in the Archean may have been
2--3 orders of magnitude more abundant than today \cite{pavlov2000},
possibly occasionally forming a Titan-like atmospheric organic haze
\cite{trainer2006,zerkle2012,arney2016}. Similar to the cloud decks of
Venus, such hazes may make characterization of the surface
environments of exoplanets challenging, especially for transit
transmission observations \cite<e.g.>{gao2020b}. Because of the long
path length slant viewing geometry inherent to transit transmission
observations, even hazes that are transparent to the surface in the
shorter path lengths relevant to direct imaging can become opaque at
elevated altitudes in transit observations \cite<e.g.>{fortney2005c}.

The start of the Proterozoic (2.3 billion years ago to 541 million
years ago) marked the rise of an oxygenated atmosphere, irreversibly
altering the redox state of the atmosphere, although oxygen abundance
during the middle Proterozoic may only have been present at low
abundances \cite[e.g. 0.1\% of the modern atmospheric
  level,]{planavsky2014b}. The rise of oxygen also meant the rise of
its photochemical byproduct, the UV-blocking ozone layer, with
profound implications for the surface habitability of our
planet. Understanding the chemical, geological, and even biological
interplay between Earth's volatile inventories and its secular
atmospheric composition, therefore, represents an important path
toward ensuring accurate interpretation of measurements of exoplanet
atmospheres and assessments of their prospective habitability.  Other
important issues relevant to exoplanetary science include:
\begin{enumerate}
\item How long did Earth's magma ocean period last, what was the
  nature of the earliest crust on the planet, and how long did it take
  for the oceans to form \cite<e.g.>{katyal2019,monteux2020}?
\item When did life originate and evolve on Earth
  \cite<e.g.>{mojzsis1996,dodd2017}? How have abiotic factors
  including (but not limited to) petrology and degassing at the ocean
  floor contributed to a changing atmospheric composition
  \cite<e.g.>{lyons2014a}?
\item How important was the Moon-forming collision for the interior
  structure and composition of Earth and the subsequent evolution of
  life here \cite<e.g.>{canup2012}?
\item What is the role of volatiles (e.g., liquid water) in
  continental plate subduction and the carbon cycle
  \cite<e.g.>{regenauerlieb2001,bercovici2003a,vanderlee2008}, and how
  critical is this process for the sustained habitability of
  terrestrial exoplanets
  \cite<e.g.>{valencia2007c,lammer2009a,noack2014b}?
\item How has the composition of the Earth's atmosphere changed with
  time due to the influence of biology \cite<e.g.>{reinhard2017}?
\item How has the depth of the oceans, and the amount of continental
  freeboard, changed through time \cite<e.g.>{korenaga2017b}, and how
  do the interaction of continents and ocean depth influence planetary
  habitability \cite<e.g.>{cowan2014,glaser2020a}?
\item What aspects of life's impact on Earth's current and past
  environments are potentially detectable on an exoplanet
  \cite<e.g.>{sagan1993d,kaltenegger2007,robinson2014c,rugheimer2015a,schwieterman2018}?
\end{enumerate}

In the context of exoplanets, Earth is invaluable as the planet we
have by far the most data for, and as the only known planet with a
biosphere. Understanding which processes are -- or are not --
necessary for habitability and the origin of life on our own planet
will help us understand the potential for habitability and life on
worlds with different histories and characteristics. Our modeling
efforts of habitable exoplanets often starts with planets analogous to
Earth. This is for good reason: grounding and validating these models
in the characteristics of our planet is a necessary step to ensure
their accuracy before they can be extended to exoplanets. Further,
understanding the diversity of ways a planet can maintain habitability
over long time periods can, together with remote sensing of our own
planet “as an exoplanet”, provide a useful starting point for future
interpretations of potentially habitable exoplanets. Exoplanet science
is rapidly moving towards a regime where observations and
characterization of such potentially habitable worlds will be
possible, and Earth is the best starting point we ever will have for
interpreting the data we obtain from these distant worlds.


\subsection{Mars}
\label{mars}

The surface of Mars today boasts a variety of features we recognize
from Earth and other planets, including impact craters, tectonic
structures, and volcanoes and their products (including the largest
such examples in the Solar System) \cite<e.g.>{werner2009b}. But
also preserved on this rocky world is evidence for a
different, early Mars, when $\sim$3.5--4.0 billion years ago liquid
water carved valley networks
\cite<e.g.>{ehlmann2014,grotzinger2014b,graugalofre2020a}, and
atmospheric pressure was much higher than the 6 mbar today
\cite{carr2012}. An early dynamo indelibly marked the ancient crust
\cite{acuna1999}, dying before the valley networks formed
\cite<e.g.>{lillis2008d}.

In terms of exoplanetary science, Mars represents a case study for a
world that was once more geologically active than today, and perhaps
even once habitable, that underwent a major change in internal and
surface properties as its interior cooled and its atmosphere was lost
to space \cite<e.g.>{ehlmann2016b}. It is possible that Mars is at or
near the minimum size of a rocky world that makes this transition,
since widespread geological activity ended much earlier on smaller
Mercury and the Moon, but continues to the present on larger Earth and
(probably) Venus \cite<e.g.>[and references therein]{byrne2020}. Mars
with its tenuous atmosphere has been considered in the context of
exoplanet atmospheric escape \cite{brain2016,dong2018b}. This is
particularly important because many exoplanets orbit very closely to
their stars and/or their stars exhibit high levels of activity that
may be sufficient to strip significant fractions of an atmosphere. A
modeling study of Mars atmospheric escape over geologic time,
validated by MAVEN observations, has suggested that 100 bars of
atmosphere can be lost from a Mars-like exoplanet orbiting in the HZ
of a M-dwarf star over 4 billion years \cite{dong2018b}. Other
important science questions for understanding Mars in the context of
exoplanetary science include:
\begin{enumerate}
\item What is the current internal structure and activity of Mars
  \cite<e.g.>{banerdt2020}, and how does that activity relate to
  surface geology \cite<e.g.>{giardini2020}, as a sub-Earth-size rocky
  world that is several billion years old?
\item To what extent was early Mars really "warm and wet" in contrast
  to the cold and dry planet we see today
  \cite<e.g.>{ramirez2018b,graugalofre2020a}?
\item How do the Red Planet's histories of dynamo generation,
  atmospheric loss, and habitability interrelate, and what we can
  learn from them for planetary habitability in general
  \cite<e.g.>{kite2019a}?
\item What is the lower size limit for sustained planetary
  habitability \cite<e.g.>{ehlmann2016b}?
\end{enumerate}

In the exoplanetary context, Mars offers us a fascinating insight into
the kind of planet that could, potentially, be mistaken for an
'Earth-like', habitable world. It offers a cautionary tale -- showing
us how planets can evolve from being eminently habitable (as it seems
was the case for the warm, wet, young Mars) to one whose habitability
is, at best, borderline or questionable
\cite<e.g.>{bishop2018,ramirez2020a}. Mars also stands testament to
the vagaries of the chaotic and violent latter stages of planet
formation, with some studies suggesting that its relatively small size
(compared to Earth and Venus) is the result of significant collisional
attrition, and others arguing that it might be more representative of
the oligarchs that precluded the final collisional growth of the two
largest terrestrial planets
\cite<e.g.>{izidoro2015c,brasser2017b,bromley2017}. The chaoticity of
Mars' obliquity, and the variability in its orbit, combine to give the
planet a far greater level of climatic variability than is seen for
Venus or the Earth
\cite<e.g.>{ward1973b,jakosky1995a,mischna2013}. Once again, this
behaviour can act as an illustration of the different factors that
could render a given 'potentially habitable exoplanet' more, or less,
suitable for life to develop and thrive - and as such, will help guide
our efforts to select the best exoplanets to target in the search for
life beyond the Solar system. Finally, studies of the Martian interior
give us an insight to a planet with failed plate tectonics - offering
ground truth for studies that model the range of planetary outcomes
for which such tectonic activity is feasible.


\section{The Giant Planets}
\label{giant}

\begin{figure*}
  \begin{center}
      \includegraphics[width=\textwidth]{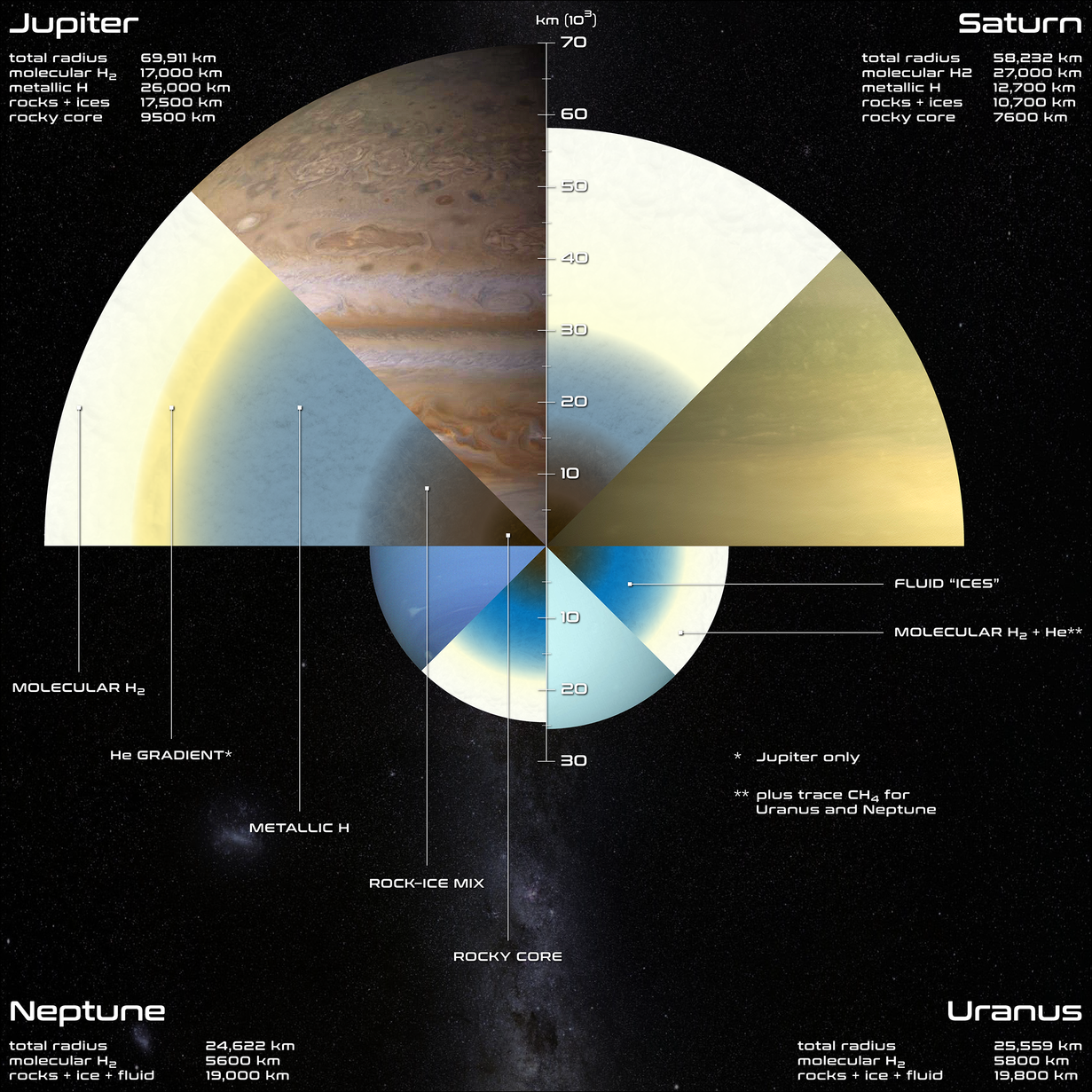}
  \end{center}
  \caption{Schematic cross sections of the four giant planets of the
    Solar System, showing the major internal components. All cross
    sections are to scale, but the thickness for each component layer
    are only approximate \cite{spiegel2014}. Those layer thicknesses
    are shown to the nearest 100~km for schematic purposes, but we
    emphasize that the interior structure of these words is not known
    to that level of precision. For this illustration, fluid "ices"
    are shown within Uranus and Neptune; other interior models are
    possible with available geophysical and spectroscopic data for
    these planets. Even so, note the substantial differences between
    the interiors of the "gas giants," Jupiter and Saturn, and those
    of the "ice giants," Uranus and Neptune.}
  \label{fig:giant}
\end{figure*}

The giant planets of the Solar System have long been the subject of
fascination and scientific investigation. Similarly, many of the
earliest confirmed exoplanets were also giant, owing largely to the
biases associated with most methods of exoplanet detection
\cite{fischer2014b}. Giant planets also hold most of the planetary
mass and angular momentum of their respective planetary systems,
making them key players in determining the final architectures of
planetary systems generally
\cite<e.g.>{morbidelli2007b,childs2019,kane2020b}. Schematics of the
interiors of the Solar System's giant planets are shown in
Figure~\ref{fig:giant}, and indicate why these worlds are termed "gas
giants" and "ice giants". Most of Jupiter and Saturn consist of some
form of H and He, including metallic hydrogen, with traces of heavier
gases and possibly even some rock and ice at their centers. Uranus and
Neptune have much higher abundances of volatiles like water and
ammonia than the gas giants, but do not have internal pressures
sufficient to generate metallic hydrogen. Instead, these ice giants
may feature sub-cloud "oceans" of a slushy mix of mainly water and
ammonia ices \cite<e.g.>{wiktorowicz2007}.
 
The presence of giant planets has often been suggested to influence
the habitability of terrestrial planets within the same system
\cite{georgakarakos2018,sanchez2018}. Such influence could include
offering shielding from an impact regime that would otherwise render
the planet sterile \cite<e.g.>{quintana2016}. However, several studies
have revealed that the situation is likely far more complex
\cite<e.g.>{horner2008a,horner2009a,horner2012b,lewis2013c,grazier2016,grazier2018},
with giant planets acting as a source of potential impact threat. Yet
such a role may be advantageous, as several models of the origin of
Earth's water invoke an exogenous source, requiring the migration of
giant planets to deliver volatiles during Earth's youth (see
Section~\ref{minor}). Thus, the effects of impacts can have positive
(volatile delivery) and negative (extinction events) consequences for
terrestrial planets. At the same time, the influence of giant planets,
or other significant perturbers, on the system orbital evolution may
play an important role in shaping the climatic variability of
potentially habitable worlds by influencing their Milankovitch cycles
\cite<e.g.>{deitrick2018a,deitrick2018b,horner2020a,kane2020e,wolf2020}.

Here, we address separately the two major classes of Solar System
giant planet, the gas giants (Section~\ref{gasgiants}) and the ice
giants (Section~\ref{icegiants}) followed by a summary of open
questions for giant planets (Section~\ref{giantquestions}). The
exploration of both groups is key to furthering our understanding of
exoplanetary systems.


\subsection{Gas Giants}
\label{gasgiants}

Numerous spacecraft have visited the mighty gas giants, Jupiter and
Saturn. Missions such as {\it Voyager 1} and {\it 2}, {\it Galileo},
{\it Cassini-Hyugens}, and {\it Juno} have directly explored the
interiors, atmospheres, magnetospheres, rings and satellites of these
worlds, and have discovered immense complexity. Rather than being a
centrally condensed planet with distinct internal layers, Jupiter
likely has a diluted, silicate-rich core that may extend to a
substantial fraction of its radius \cite{wahl2017b}. Saturn may also
have compositional gradients \cite{iess2019}, despite evidence for a
rocky core with mass $\sim$15~$M_\oplus$
\cite{movshovitz2020}. Understanding the relation between interior
complexity and bulk composition for gas giants is pivotal in
interpreting exoplanet observations, which only yield the latter of
properties. For example, \citeA{thorngren2016} found that the bulk
heavy element masses of their sample of 47 giant planets only modestly
changed when assuming different equations of state.

Exploration of the atmospheres of Jupiter and Saturn has also provide
invaluable information for the interpretation of direct or indirect
observations of exoplanet atmospheres. The Galileo entry probe
measured the abundances of various gas species in Jupiter's atmosphere
including He, which was necessary to interpret the thermal evolution
of the planet \cite<e.g.>{vonzahn1998c}. The probe measured
abundances of heavy elements C, N, S, and P and the heavy noble gases
Ar, Kr, and Xe that were enhanced relative to solar by a factor of
2--4 \cite{mahaffy2000,wong2004b}, measurements that are critical for
understanding Jupiter's formation
\cite{owen1999b,gautier2001,mousis2019}. Remote sensing measurements
suggest that Saturn is enriched in C \cite{lellouch2001}, S
\cite{briggs1989b}, and P \cite{fletcher2009e} by a factor of 10--12
relative to solar but N is only enriched by a factor of $\sim$2
\cite{fletcher2011d} which could have important implications for the
formation of Saturn \cite{mandt2020a}. The {\it Galileo} probe showed
that Jupiter is depleted in He and Ne because helium likely
precipitates as droplets in the deep atmosphere \cite{stevenson1977e}
with neon being sequestered in these droplets
\cite{wilson2010}. Saturn is also depleted in He by the same process,
but no Ne measurement is available because noble gases heavier than He
can only be measured by a probe \cite{mousis2014d}. The
\textit{Galileo} probe also found a depletion in oxygen
\cite{mahaffy2000}, frequently interpreted to mean that the probe
sampled a meteorologically anomalous region of Jupiter's atmosphere
\cite{orton1998}. The {\it Juno} mission, instead discovered deep
currents that circulate ammonia and water around the planet
\cite{bolton2017e,li2017,li2020}. The in-situ measurements from the
      {\it Galileo} probe have been most valuable for providing tools
      that can be used to determine how giant planets formed and
      evolved since formation \cite<e.g.>[and references
        therein]{mandt2020b}. In-situ observations of the giant
      planets are important for exploring giant exoplanet formation
      and evolution. In particular, comparisons of the relative
      abundances of heavy elements (e.g. C/N) provide a direct
      comparison for Solar System analog exoplanets -- or at the very
      least a necessary starting point for models of more exotic
      (e.g., highly irradiated) giant exoplanets for which no direct
      analog exists in the Solar System.


\subsection{Ice Giants}
\label{icegiants}

The so-called ice giants, Uranus and Neptune, are particularly notable
in regards to exoplanets because they represent examples of what seems
to be the most common type of exoplanet yet detected. By size, the
majority of exoplanets within 100 days orbital period have radii
between that of Earth and those of Uranus and Neptune
\cite<e.g.>{fulton2018b}. Importantly, this class of exoplanet does
not occur among the ranks of the Solar System, and there appear to be
significant differences between their composition and formation
compared with the Solar System ice giants
\cite{owen2017c,lee2019}. Even so, there remain many questions as to
how mass, radius, and bulk density affect or are related to the
interior structure of the ice giant planets. Further challenges in
modeling the ice giant interiors relate to the non-dipolar and
non-axisymmetric nature of their magnetic fields
\cite{ruzmaikin1991,nellis2015a}, particularly in relation to the
dynamics and chemistry of their upper and deep atmospheres
\cite{stanley2004,redmer2011}. Uranus and Neptune (and the Earth)
bookend a much larger demographic that includes rocky and gaseous
planets. Coming from the large end of this transitional regime, Uranus
and Neptune are best windows we have to most common class of exoplanet
yet known \cite{kane2011d,wakeford2020b}.

Uranus and Neptune have only been the subjects of flybys by the
\textit{Voyager 2} spacecraft and of Earth-based observation
\cite<e.g.>{smith1986,tyler1986,lindal1987,smith1989,lindal1992,fletcher2014a}. They
have not yet been explored with orbiters or entry probes, despite
compelling planetary and exoplanetary motivation
\cite<e.g.>{atreya2020,mousis2020b,wakeford2020b,mandt2020a,mandt2020b}.
Such a mission could provide much needed context for to the growing
number of mass, radius, and atmospheric abundance measurements being
acquired for exoplanets. One aspect of the ice giants fundamental to
understanding exoplanets is their interior structures. A three-layer
model of rock, ice, and H--He gas is often employed in studies of
these worlds, but yields results that are at odds with the expected
interior ice--rock ratios of Uranus and Neptune
\cite{nettelmann2013a}. Furthermore, the formation and migration of
Uranus and Neptune are key points for comparison with exoplanets that
may have either formed closer to their host star through different
mechanisms, or experienced substantially different migration
histories. Had Uranus and Neptune formed via core accretion
\cite{mizuno1980a,pollack1996} at their current orbits, the timescale
of their formation would be longer than the lifetime of the protosolar
nebula \cite{pollack1996}. This complication can be overcome with
various assumptions involving planetary migration
\cite<e.g.>{dodsonrobinson2010}, but accurately reproducing the
measured properties of Uranus and Neptune requires very finely tuned
conditions \cite{helled2014a}. On one hand, planetary migration
likely accounts for the wide diversity of exoplanets similar in size
and mass to Uranus and Neptune since, as described by
\citeA{helled2014a}, the mass and solid-to-gas ratios are sensitive to
the birth environments of the planets. On the other hand, this freedom
in parameter space necessitates that specific information for Uranus
and Neptune be obtained with which to constrain prospective formation
scenarios; that information can be attained by future exploration of
the Solar System's ice giants \cite{mousis2020b,mandt2020b}.


\subsection{Outstanding Questions}
\label{giantquestions}

The giant planets are important analogs for a large number of known
exoplanets. Exploring the four giant planets in our solar system
allows us to better understand the formation and evolution of both gas
and ice giants as well as the opportunity to explore how giant planets
migrate after formation. Observations that have provided important
advancements in understanding the giant planets as exoplanet analogs
include the \textit{Galileo} probe measurements
\cite<e.g.>{atkinson1998,folkner1998,niemann1998c,sromovsky1998,vonzahn1998c,wong2004b}
and \textit{Juno} and \textit{Cassini} gravity measurements
\cite<e.g.>{folkner2017,kaspi2017,moore2017c,wahl2017b,guillot2018,iess2019,movshovitz2020,stevenson2020b,buccino2020a,duer2020}.
Some of the most significant remaining science questions for
understanding the giant planets in the context of exoplanetary science
include:
\begin{enumerate}
\item How did the giant planets in our Solar System form and how did
  they evolve internally after formation? How has this affected the
  composition of the observable atmosphere
  \cite<e.g.>{stevenson1977d,mizuno1980a,bodenheimer1986,pollack1996,fortney2010c,nettelmann2013a,helled2014a,dalba2015,koskinen2018,atreya2020}?
\item Did the giant planets migrate after formation and how did this
  migration impact the architecture of the Solar System
  \cite<e.g.>{goldreich1980b,lin1986c,ida2004a,tsiganis2005b,gomes2005b,morbidelli2010b,walsh2011a,walsh2011c,clement2019b}?
\item Why are the magnetic fields of the ice giants so drastically
  different from any other magnetic fields in our Solar System and
  what does that mean for the interiors of the ice giants
  \cite<e.g.>{warwick1986,warwick1989b,connerney1993a,stanley2004,stanley2006,jacobson2009,helled2010b,jacobson2014f,helled2020a}?
\item How would the phase curves of the four giant planets compare to
  what future direct imaging missions will observe in exoplanetary
  systems
  \cite<e.g.>{madhusudhan2012a,mayorga2016,mendikoa2017,macdonald2018a}?
\end{enumerate}

Understanding how our own giant planets formed and evolved and how
this has impacted the composition of the atmosphere is important for
interpreting measurements of giant planet atmosphere composition and
connecting these measurements to the history of that planetary
system. Furthermore, we are only beginning to understand the role that
the migration of giant planets plays in the architecture of a
planetary system and the delivery of volatiles, particularly water, to
planets that formed inside the water ice line. Understanding magnetic
fields is fundamental to interpreting the near space environment of an
exoplanet and how its star influences its atmosphere. The least
understood and most surprising magnetospheres in our solar system are
those of Uranus and Neptune, which demonstrate how little we truly
understand about giant planet magnetospheres. Until we have a better
understanding of them, we will be limited in what we can learn about
exoplanets. Finally, observations of the phase curves of all four
giant planets (including polar perspectives for comparison to face-on,
directly imaged systems) is critical for interpreting the phase curves
that we will eventually measure for exoplanets.


\section{Icy Moons}
\label{moons}

\begin{figure*}
  \begin{center}
      \includegraphics[width=\textwidth]{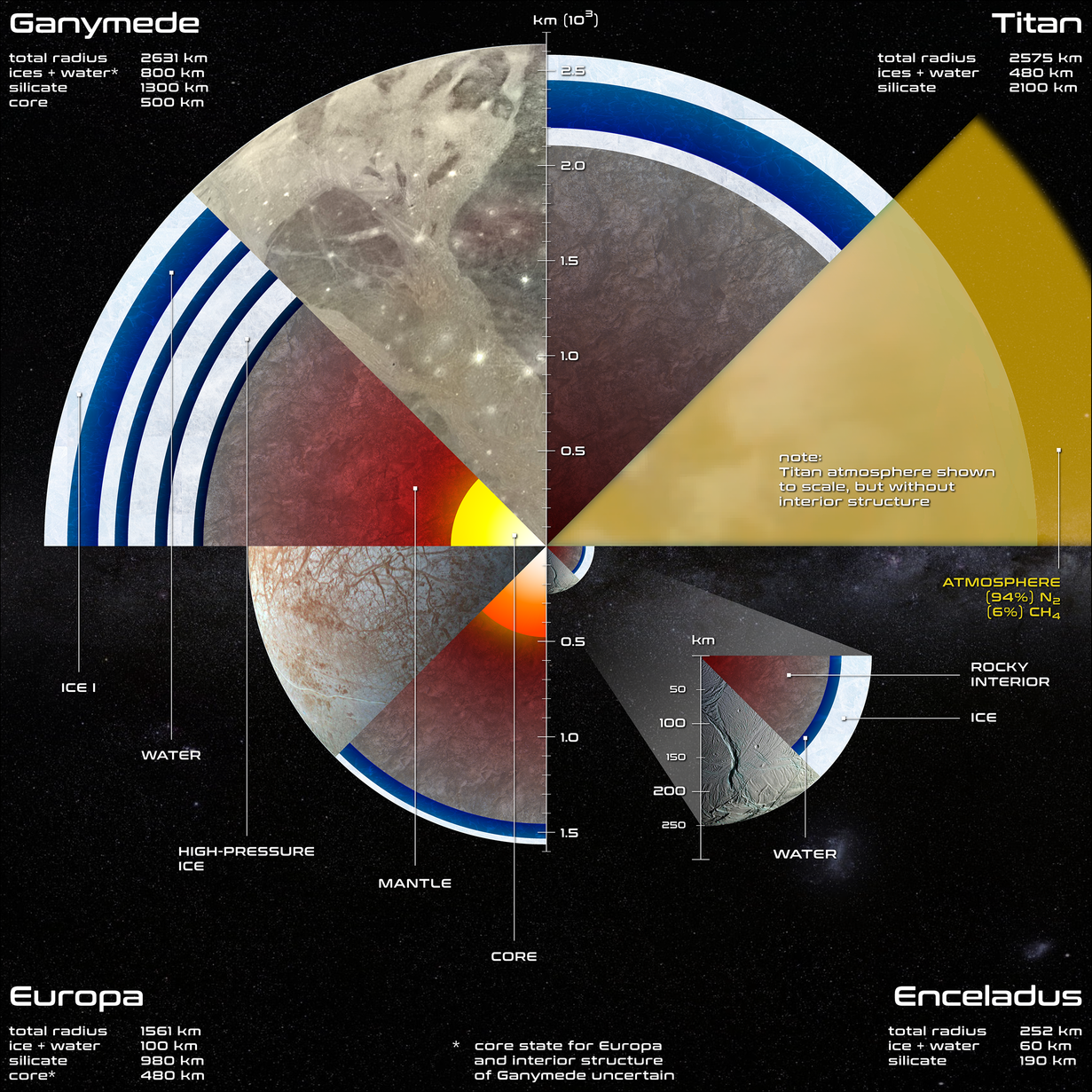}
  \end{center}
  \caption{Schematic cross sections of four of the Solar System's
    major icy moons, showing the major internal components. All cross
    sections are to scale, but the depths to each component layer
      are only approximate (based on the interior structure models of
      \citeA{vance2018b}). Depths are given to the nearest 10 km, such
      that aggregate depths may not match known planetary radii
      values. For this illustration, the interior of Ganymede is shown
      with interleaved oceans and (high-pressure) ice layers, but
      other internal arrangements are possible.}
  \label{fig:moons}
\end{figure*}

Given the prevalence of satellites within the Solar System,
substantial effort is being devoted to the search for moons orbiting
exoplanets
\cite<e.g.>{hinkel2013b,kipping2013d,heller2014c,hill2018}. Furthermore,
formation of regular moons, such as those in the Galilean system, may
serve as analogs of compact exoplanetary systems in terms of their
formation and architectures
\cite{kane2013e,makarov2018,dobos2019,batygin2020b}. However,
there are numerous questions that remain regarding the wide array of
moons in the Solar System, including their geology and, in some cases,
atmospheres. The icy satellites of the giant planets may serve as
small-scale analogs for low-mass, water-rich exoplanets, i.e.,
so-called "ocean planets." Ocean planets are a class of terrestrial
exoplanets with substantial water layers that may be common throughout
the galaxy \cite{leger2004,raymond2006d,ehrenreich2007a,sotin2007b},
and which are likely to have H$_2$O contents at least an order of
magnitude greater than Earth's $\sim$0.1\% H$_2$O content. Ocean
planets may exist in one of a variety of climactic states including,
ice-free, partially ice covered, and completely frozen
\cite{tajika2008,quick2020b}; those with highly eccentric orbits may
well also possess substantial amounts of internal energy owing to
tidal heating from their host stars. As liquid water and energy are
both necessary ingredients for life, ocean planets represent
prospective habitable environments beyond typical Earth-like
environments in the traditional HZ \cite{glaser2020a}. Indeed, even
those such worlds that are mostly ice-covered may have considerable
regions of unfrozen land near their equators or small, equatorial
regions of salt-rich water where life could flourish
\cite{delgenio2019a,paradise2019,olson2020}.

Shown in Figure~\ref{fig:moons} is a representation for the interior
structures of the icy moons of our outer Solar System's giant planets,
highlighting the diversity of internal structures. Studying the
interiors, tidal properties, and evolution of these moons may provide
similar key insights into the properties of ocean planets
\cite{ehrenreich2007a,sotin2007b,henning2014,vance2007,noack2016,luger2017b,yang2017,barr2018,journaux2020b}. Owing
to their similar internal structures, geophysical processes operating
on ocean planets with ice-covered surfaces may be similar to
geophysical processes operating on the moons of the giant planets, and
may include ice tectonics \cite{fu2010,levi2014,hurford2020}, and
cryovolcanism \cite{levi2013,quick2020b}. Although the specular
reflection of starlight, or "glint", on the surfaces of ocean-covered
planets will make them fairly easy to detect at visible and near-IR
wavelengths
\cite{williams2008,robinson2010,visser2015,lustigyaeger2018}, the high
albedos of ocean planets with ice-covered surfaces will make them far
more detectable than rocky planets in reflected light
\cite{wolf2017b}. Many ocean planets may resemble Saturn's largest
moon Titan (Figure~\ref{fig:moons}), where the presence of a dense
atmosphere allows for the maintenance of liquid at its surface
\cite{lora2015a}. With its active methane cycle
\cite{turtle2011b,dalba2012b,horst2017,levi2019}, Earth-like
shorelines \cite{lunine2009}, diverse geological processes
\cite{jaumann2009}, and the potential for prebiotic chemistry
\cite{neish2009,he2014b}, Titan serves as an analog for ocean planets
that are similar to Earth in nature. Haze in the atmospheres of
Titan-like exoplanets could be detected by next-generation space
telescopes \cite{robinson2014d,checlair2016,lora2018}, thereby
revealing the atmospheric compositions of numerous ocean planets.


\section{Minor Planets}
\label{minor}

Our Solar System informs our understanding of volatile distribution
and planet migration, especially from careful study of its minor
bodies: asteroids and Edgeworth-Kuiper Belt Objects (KBOs). Asteroids,
and the meteorites that sample them, map out the distribution of water
in our protoplanetary disk. For excellent reviews of different
meteorite types, and their connections to asteroids, we refer the
reader to \citeA{weisberg2006} and \citeA{demeo2015a}. Asteroids are
categorized by their reflectance spectra. Three main types are: E-type
asteroids, at $\approx$1.9--2.1~AU; S-type asteroids, at
$\approx$2.1--2.7~AU and beyond; and C-type asteroids, at 2.7--3.5~AU,
though some lie interior to these distances
\cite{gradie1982c,binzel2019}. Meteorites from unmelted asteroids are
termed chondrites, and are categorized according to major elemental
distributions as well as isotopic anomalies \cite{weisberg2006}, and
are presumed to come from parent bodies sampling a variety of
heliocentric distances in the protoplanetary disk. The different
asteroids are spectrally associated with the three main types of
chondrites: enstatite chondrites (ECs), associated with E-type
asteroids; ordinary chondrites (OCs), with S-type asteroids; and
carbonaceous chondrites (CCs), with C-type asteroids
\cite{gaffey1993b,binzel2019}. Generally, CCs are the most
volatile-rich, with abundant hydrated phases equivalent to a few wt\%
H$_2$O in CO and CV CCs, up to 13wt\% in CM and CI CCs
\cite{alexander2013}. ECs are the least volatile rich, with no
hydrated phases, and sulfides and other reduced minerals that would
have been destroyed by water on the parent body. OCs have
$\sim$0.1--1wt\% H$_2$O, indicating that they formed outside the
H$_2$O snow line, but in a region with much lower water ice abundance
than where most CCs formed. C-type asteroids appear to have been
scattered into their present orbits from beyond Jupiter, but E-type
and S-type asteroids seem to have formed in place
\cite{walsh2011c}. This places the snow line between where ECs and OCs
may have formed, i.e., at 2~AU, at the time they formed, about 2~Myr
\cite{desch2018}.

Earth formed inside the snow line \cite{wetherill1990a,raymond2004a},
but acquired water by accreting materials from beyond the snow
line. From an elemental and isotopic perspective, Earth resembles a
mix of about 71\% ECs, 24\% OCs, and 5\% CCs, of type CO or CV
\cite{dauphas2017}. Assuming the 29\% of its mass that is OCs and CCs
had $\sim$1wt\% H$_2$O, Earth would have roughly 0.003~$M_\oplus$ of
H$_2$O (or about 12 oceans' worth of water), a good match to the
inferred amounts of hydrogen in Earth's core, mantle, and surface,
equivalent to about 0.002~$M_\oplus$ \cite{wu2018}. But Earth could
have had much more water if it had accreted a larger fraction of
material from beyond the snow line, or especially if the material just
beyond the snow line was more water-rich. OCs, despite forming in a
region cold enough for water ice to condense, ended up containing only
0.1--1wt\% water, instead of the few to 13wt\% H$_2$O seen in
CCs. \citeA{morbidelli2016a} explained this in terms of a ``fossil
snow line", in which Jupiter formed and grew large enough to open a
gap in the disk (more precisely, create a pressure maximum in the disk
outside its orbit), while the snow line was exterior to Jupiter;
later, even as accretion waned, the disk cooled and the snow line
formally moved inward, ice could not follow because most of it was
bound in large (cm-sized) particles that remained trapped in the
pressure maximum. Similar ideas were invoked by \citeA{kruijer2017a}
to explain the isotopic dichotomy of the Solar System, and by
\citeA{desch2018} to explain the distribution of calcium-rich,
aluminum-rich inclusions (CAIs) in chondrites. In both models, Jupiter
must grow to 20--30~$M_\oplus$, to create a pressure maximum by about
0.4--0.9~Myr. The detailed disk model of \citeA{desch2018}, which
includes accretion heating and is tailored to fit multiple constraints
from 18 different meteorite types, predicts the snow line at 2~Myr
should have been at 2~AU, and conforms to the fossil snow line model
of \citeA{morbidelli2016a}.

In our Solar System, the snow line in the protoplanetary disk stage
was at 2 AU \cite{rubie2015}, just beyond the (future) HZ at about 1
AU, leading Earth to be a habitable, but relatively volatile-poor
(0.025wt\% surface H$_2$O) planet. The relative positions of the HZ
and snow line would be different in exoplanetary systems around other
stars, since they have different dependencies on the luminosity and
effective temperature of the host star. For example, the HZ planets
orbiting the late M star TRAPPIST-1 may be more volatile-rich than the
Earth. The masses and radii of planets f and g, orbiting in the HZ of
TRAPPIST-1, seem to demand $\approx 50$wt\% H$_2$O
\cite{unterborn2018a}. These planets likely formed much farther
(perhaps 4 times farther) from their host star, beyond the snow line,
and then migrated inward \cite{unterborn2018a}. Migration is supported
by the fact that all the planets are nearly in mean motion resonances,
with period ratios supportive of convergent migration
\cite{steffen2015}. For the Solar System, both the inward and outward
migration of the giant planets is strongly signposted by the
distribution of the Solar system's small body populations -- with such
migration having sculpted the Asteroid and Edgeworth-Kuiper belts
\cite<e.g.>{hahn2005b,levison2008a,minton2009,morbidelli2010a,demeo2014b},
and resulted in the capture of the Jovian and Neptunian Trojans
\cite<e.g.>{morbidelli2005,lykawka2009b,lykawka2010b,pirani2019a} and
the Plutinos \cite<e.g.>{malhotra1993b,malhotra1995b}

While the rocky planets in our Solar System do not appear to have
migrated, minor bodies strongly indicate that the giant planets
migrated. The asteroid belt today has only about 0.1\% of the rocky
mass that probably existed between 2--3~AU during the protoplanetary
disk phase, and S-type and C-type asteroids are commingled in this
region, both facts possibly explained by Jupiter's migration
\cite{minton2009,walsh2012c}. Meteorites appear to record several
large impacts in the Solar System around 5~Myr, including the impact
of the ureilite parent body \cite{amelin2015} and the CH/CB parent
body \cite{krot2008a}. Presumably the outward migration of Jupiter in
either model would have depleted the asteroid belt and scattered the
C-type asteroids into the asteroid belt.  Large-scale migration of
Jupiter is not necessarily demanded to mix C-type and S-type asteroids
in the asteroid belt \cite<e.g.>{raymond2017b}, but it is a common
feature of dynamical models of the early Solar System
\cite{clement2019a,tsiganis2005b}. The models of \citeA{walsh2011c}
and \citeA{desch2018} both rely on the rapid ($\sim 10^5$ yr)
migration of Jupiter and Saturn while $< 10$~$M_\oplus$, then slower
($> 10^6$ yr) migration after growing to masses large enough to open a
gap in the disk, as commonly theorized for growing planets
\cite<e.g.>{bitsch2015b}. Thus, studies of Solar System minor bodies
reveal important lessons for understanding rocky exoplanets and the
role of Jupiter analogs, and suggest that the majority of systems may
have more volatile-rich rocky exoplanets, but might be characterized
by even more orbital migration.


\section{Exoplanets and Observables}
\label{exoplanets}

\begin{figure*}
  \begin{center}
      \includegraphics[width=\textwidth]{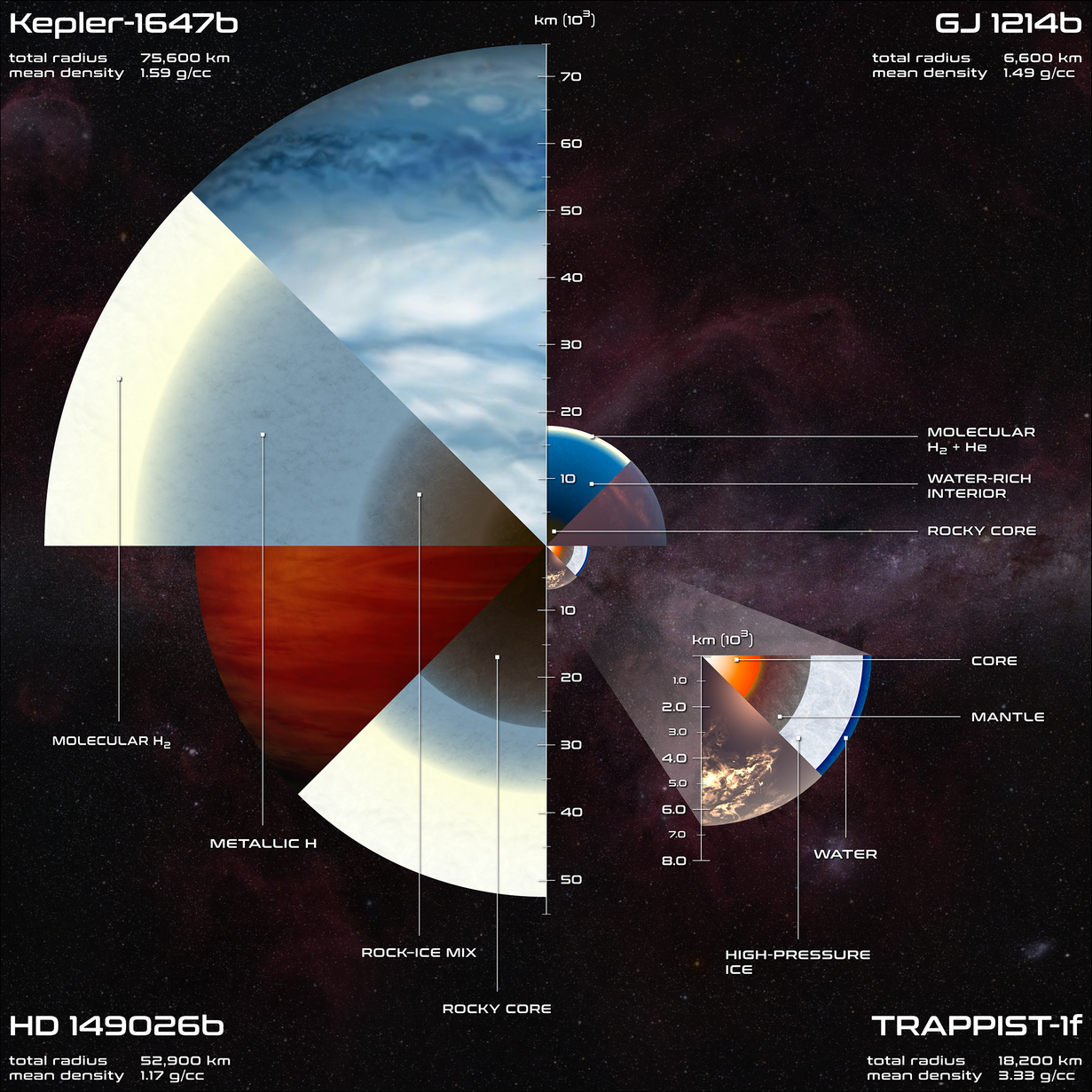}
  \end{center}
  \caption{Schematic cross sections of four selected exoplanets that
    span a broad range of sizes and predicted interior structures and
    compositions: Kepler-1647b, HD~149026b, GJ~1214b, and
    TRAPPIST-1f. All cross sections are qualitatively to scale, but
    the structure and composition of these interiors, and of these
    planets' atmospheres, are uncertain and are shown here for
    illustrative purposes only. Kepler-1647b and HD~149026b are
    analogous to the Solar System gas giants and TRAPPIST-1f may be
    analagous to Venus and/or Earth. GJ~1214b is likely a water-rich
    mini-Neptune that represents a size regime between that of Earth
    and Neptune. The radii and density values are from the NASA
    Exoplanet Archive \cite{akeson2013}.}
  \label{fig:exoplanets}
\end{figure*}

While the Solar System is our best studied example of a planetary
system, observations of exoplanets have expanded our horizons to
reveal planets and planetary system architectures that are unknown in
our home system. These alien systems have helped reveal key planetary
processes that refine our understanding of how our own planets and
planetary system might have formed and evolved. In particular, the
discovery of planetary types not found in the Solar System, including
hot Jupiters, sub-Neptunes and volatile-rich terrestrials has helped
us better understand fundamental processes such as atmospheric loss
and planetary migration, that have also sculpted our own
planets. Shown in Figure~\ref{fig:exoplanets} are four examples of
exoplanets shown to scale but spanning a broad range of size, density,
interiors, and atmospheres. These include Kepler-1647b, a Jupiter
analog orbiting a binary star \cite{kostov2016a}, HD~149026b, a dense,
giant planet with large core \cite{sato2005c}, GJ~1214b, a water-rich
mini-Neptune \cite{charbonneau2009,rogers2010b}, and TRAPPIST-1f, a
potential ocean-planet in the HZ
\cite{gillon2017a,unterborn2018a}. Though super-Earth and mini-Neptune
planets are not represented in the Solar System, GJ~1214b and
TRAPPIST-1f may represent two examples of ocean worlds, possibly
similar to icy moons of the Solar System, described in
Section~\ref{moons}.

The observing techniques used to study exoplanets are currently most
sensitive to planets that are close to (via transit and radial
velocity), or very far from their stars (via direct imaging and
astrometry) and the region in between where most of the Solar System
planets would reside is currently relatively inaccessible. This lack
of overlap makes it more difficult to place our Solar System in its
true cosmic context, but nonetheless provides an excellent opportunity
to study planetary systems very unlike our own that challenge long
held concepts. These include systems, some orbiting G dwarfs like our
Sun, where multiple planets are found within the equivalent of the
orbit of Mercury \cite{lissauer2011b}, speaking to the importance of
planetary migration as a fundamental system process
\cite<e.g.>{walsh2011a,gillon2017a,luger2017b,ramos2017,unterborn2018a},
and also highlighting the likely importance of gravitational
interactions and tidal heating, seen primarily in the giant planet
satellites in our system, to close-in exoplanets. Extrapolations of
the likely demographics in the not yet fully explored regions of
exoplanet systems also hint at the relative rarity of Jupiters, with
only about 10\% of solar type stars and 3\% of M dwarfs harboring
giants inside of 10~AU
\cite{zechmeister2013,wittenmyer2016c,wittenmyer2020a}. Upcoming
observations, including the Roman Space Telescope microlensing survey
will detect many more planets at similar distances to their star as
our planets, and provide the statistics needed to better understand
how common planetary Systems like the Solar System are in our galaxy
\cite{penny2019}. The direct imaging capabilities of Roman will be
able to access analogs of both Jupiter and Neptunes, providing rare
insights into the atmospheres of ice giants external to the Solar
System \cite{lacy2019}.

After over a decade of giant exoplanet characterization, the field is
on the brink of terrestrial exoplanet atmosphere observations. We have
obtained transmission spectra of a suite of hot Jupiters, revealing a
diversity of giant worlds with a range of different atmospheric
compositions, clear or cloudy atmospheres, and temperature profiles
with strong stratospheric inversions or none at all
\cite{sing2016}. Observations of exo-Neptunes are now state of the art
\cite{crossfield2017b}, revealing flat to water absorption dominated
spectra that show possible trends in composition with size, with the
smaller planets having a higher fraction of elements heavier than
hydrogen. These trends are similar to those seen in atmospheric
composition for Solar System giants. But perhaps the most exciting
advances of all are the first observational constraints on
terrestrial-sized worlds. Transmission spectroscopy of the Earth-sized
TRAPPIST-1 planets rule out cloud-free, H$_2$-dominated atmospheres
\cite{dewit2016b,dewit2018,wakeford2019}, and a combination of
laboratory work and modeling suggests that the atmospheres of these
worlds, if they exist are also unlikely to be H$_2$-rich and cloudy
\cite{moran2018}. These combined constraints suggest that these
terrestrial worlds may have high-molecular-weight atmospheres like the
terrestrial planets in our own system, although the data are also
consistent with no atmospheres at all. With the launch of the James
Webb Space Telescope (JWST), observations of the TRAPPIST-1 system and
other nearby terrestrial worlds will have the capability to detect the
presence and composition of atmospheres
\cite{morley2017b,lincowski2018,lustigyaeger2019a,wunderlich2019},
potentially revealing past processes like atmosphere and ocean-loss
\cite{lincowski2018,lincowski2019,lustigyaeger2019b}. These
observations may also provide our first opportunity to search for
signs of life, such as CH$_4$ in combination with other biosignatures,
in the atmosphere of a terrestrial exoplanet
\cite{krissansentotton2018c,wunderlich2019}, and complement
observations from the ground with extremely large telescopes that will
search for O$_2$ using high resolution spectroscopy
\cite{lovis2017,lopezmorales2019}.

Arguably, the overarching goal of exoplanetary science is to find
reliable pathways towards accurate characterization of exoplanet
atmospheres, surfaces, and interiors. Fundamental exoplanet
observables such as mass and radius can be used to determine density,
which in turn constrains planetary bulk composition. Transmission
observations and direct imaging can reveal atmospheric composition,
and in the case of direct imaging, potentially surface composition as
well. However, correct interpretation of all of these data rely upon
models of planetary processes that are best developed and validated
using in-situ or remote-sensing observations of Solar System bodies
\cite{fujii2014}. Similarly, observations of Solar System bodies,
even very fundamental ones like phase dependent photometry of the
Jovian planets \cite{mayorga2016}, or simulated transmission
observations of Titan \cite{robinson2014d} or the Earth
\cite{macdonald2019c} can help inform planning and interpretation of
exoplanet observations, and help us train predictive (forward) and
retrieval (inverse) models for exoplanets.

Both the exoplanet and Solar System planetary communities are moving
towards a more systems and processed-based approach to understanding
planet formation, evolution, habitability and biosignatures. These
approaches require the synthesis of observations, theory and
laboratory work from multiple disciplines, and it is very clear that
the two communities can benefit greatly from the knowledge and
perspectives provided by both of their fields. As described above,
planetary science forms the foundation, both in terms of data and
models, from which exoplanet observables may be interpreted. In turn,
exoplanet observables provide vast numbers of exoplanets from which
demographic studies can inform the studies of planetary system
formation and evolution in general
\cite<e.g.>{clanton2016,barclay2017,nielsen2019c}. Measurements
(both direct and indirect, respectively) from planetary science and
exoplanet observables feed into the inferred properties of planetary
bodies generally and, in particular, the potential surface conditions
of a terrestrial exoplanet that may have temperate surface
conditions. Models that are well-validated on Solar System bodies,
especially Earth, will be particularly critical for the difficult task
of inferring the presence of life on an exoplanet from possible
observed biosignatures. Indeed, this may be the most challenging task
faced by planetary scientists in the coming decades when spectral
observations of potentially habitable terrestrial planets become
possible, and interdisciplinary collaborations of scientists will be
essential to its success.

The pathway forward therefore lies in identifying the key measurables
from Solar System bodies, through in situ observations with spacecraft
missions and those taken on and in orbit of Earth
\cite{robinson2011a,jiang2018}, needed to correctly interpret
exoplanet observables and infer their properties. In the near-term,
the most critical data needed from planetary science are atmospheric
measurements that can constrain composition, chemistry, dynamics and
evolutionary history, particularly for poorly understood atmospheres
such as those of Venus \cite{kane2019d}, Titan \cite{checlair2016},
and the ice giants \cite{wakeford2020b}. Beyond modeling the nature of
exoplanet atmospheres and inferring possible surface conditions lies
the complex task of modeling how the interior and surface have
previously, and are currently, interacting with the atmosphere. In
particular, a major challenge lies in distinguishing between biotic
and abiotic processes that yield gases of biological significance
(e.g., CH$_4$, CO$_2$, H$_2$O, etc.), the signatures of which can be
detected from studying atmospheric abundances
\cite{meadows2017,fujii2018,harman2018b,wogan2020}.


\section{Conclusions}
\label{conclusions}

In the current era there are two separate but complementary fields:
planetary science and exoplanetary science. Historically, the reason
for the separate pathways of the two fields resulted from exoplanet
detection primarily being a task of stellar characterization by
stellar astronomers, from which the presence of a companion of
planetary size and/or mass may be inferred, whereas planetary science
has focused on specific worlds within the Solar System. However,
discoveries of terrestrial exoplanets has provoked further discourse
between the disciplines as we strive towards a common objective of
leveraging Solar System data towards a deeper understanding of the
exoplanet observables. In time, we may come to understand planetary
bodies at the systems level, with perhaps the ultimate goal to
understand where else in the cosmos we might search for, and find,
life.

Planetary science has an exceptionally long history of ground and
space-based observations of Solar System bodies, along with robotic
exploration and in-situ analysis of atmospheres, surfaces, and
interiors. These data have provided the foundation for our fundamental
understanding of planetary processes and the signatures that those
processes produce. In this work, we have provided a brief overview for
some of the research highlights of planetary science, particularly as
these discoveries relate to exoplanets. Although we have presented
this information in categories of terrestrial planets, giant planets,
moons, and minor bodies, there is clearly substantial overlap between
these classes of objects in their formation, structure, evolution, and
interaction with each other. Additionally, we have outlined some key
questions that remain for various Solar System bodies, the answers of
which will further inform the models used in the interpretation of
exoplanet observations.

Given the increasing rate of exoplanet discoveries and the rapid
expansion of exoplanet characterization studies, the trajectory of
exoplanetary science in the years ahead is expected to require
detailed modeling of planetary atmospheres and their interaction with
surface and interior processes. As described in this work, near-term
testing of exoplanet models will rely on Solar System data, and indeed
the design of many recent Solar System exploration proposals are
incorporating science goals that specifically benefit anticipated
studies of exoplanets. Furthermore, laboratory experiments are being
conducted that provide the framework for interpreting molecular
absorption in exoplanet atmospheres
\cite{tennyson2017,horst2018b,he2020b,moran2020}. Therefore, it is
expected that the continuing reliance of exoplanetary science on Solar
System exploration will provide enormous benefits for both fields of
research as the vast data provided by the plethora of exoplanets
answers significant questions regarding the context and uniqueness of
the Solar System and, ultimately, the prevalence of life in the
universe.


\acknowledgments

This work benefited from discussions at the "Exoplanets in our
Backyard" workshop held in Houston, USA, February 5-7, 2020. This
research has made use of the Habitable Zone Gallery at hzgallery.org,
and the NASA Exoplanet Archive, which is operated by the California
Institute of Technology under contract with the National Aeronautics
and Space Administration under the Exoplanet Exploration Program. The
results reported herein benefited from collaborations and/or
information exchange within NASA's Nexus for Exoplanet System Science
(NExSS) research coordination network, which is sponsored by NASA's
Science Mission Directorate. P.D. is supported by a National Science
Foundation (NSF) Astronomy and Astrophysics Postdoctoral Fellowship
under award AST-1903811. G.A. acknowledges support from the NASA
Astrobiology Institute's Virtual Planetary Laboratory, supported by
the NASA Nexus for Exoplanet System Science (NExSS) research
coordination network Grant 80NSSC18K0829, and from the Goddard Space
Flight Center Sellers Exoplanet Environments Collaboration (SEEC),
which is funded by the NASA Planetary Science Division's Internal
Scientist Funding Model (ISFM). K.M. acknowledges support from NASA
RDAP grant 80NSSC19K1306. Data were not used, nor created for this
research.




\end{document}